# DispersioNET: Joint Inversion of Rayleigh-Wave Multimode Phase Velocity Dispersion Curves using Convolutional Neural Networks

Rohan Sharma[*1], Divakar Vashisth[2] and Bharath Shekar[3]

[1]*Department of Applied Geophysics, Indian Institute of Technology Dhanbad*, [2]*Department of Energy Science and Engineering, Stanford University*, [3]*Department of Earth Sciences, Indian Institute of Technology Bombay*

*email:* [1]*rohanlatha29@gmail.com,* [2]*divakar.vashisth98@gmail.com*, [3]*bshekar@iitb.ac.in*

## Summary

Rayleigh wave dispersion curves have been widely used in near-surface studies, and are primarily inverted for the shear wave (S-wave) velocity profiles. However, the inverse problem is ill-posed, non-unique and nonlinear. Here, we introduce DispersioNET, a deep learning model based on convolution neural networks (CNN) to perform the joint inversion of Rayleigh wave fundamental and higher order mode phase velocity dispersion curves. DispersioNET is trained and tested on both noise-free and noisy dispersion curve datasets and predicts S-wave velocity profiles that match closely with the true velocities. The architecture is agnostic to variations in S-wave velocity profiles such as increasing velocity with depth and intermediate low-velocity layers, while also ensuring that the output remains independent of the number of layers.

## Introduction

Near-surface studies have multiple applications that utilize the inversion of surface wave dispersion curves. It is used in various areas, like assessing the potential of soil liquefaction (Lin et al., 2004), passive seismic data analysis (Shekar et al., 2023), etc. Rayleigh waves can be utilised to estimate shear (S-) wave velocities for near-surface characterisation. The problem of inverting Rayleigh wave dispersion data is ill-posed, non-unique and nonlinear (Cox and Teague, 2016). Several conventional local and global optimisation methods like genetic algorithm (e.g. Moro et al., 2007) and swarm intelligence-based algorithms (e.g. Vashisth and Shekar, 2019) have been applied to tackle this problem. However, often, the inversion process is solely focused on the fundamental mode. The fundamental mode of Rayleigh waves is primarily sensitive to shallow structures, making it less effective for probing deeper geological formations, especially beneath intermediate low-velocity and high-velocity layers (Shen et al., 2016). Deeper features have a greater impact on the higher phase velocities that correspond to the frequencies of higher modes. Therefore, these higher modes assist in the delineation of deeper strata because of an increase in penetration depth (Pan et al., 2019). Consequently, uncertainties in the estimated S-wave velocity profiles could be reduced by performing a joint inversion of dispersion curves for the fundamental and higher-order modes.

Deep learning methods have gained immense popularity in the geophysical inversion domain because of their ability to approximate the complex nonlinear functional mapping between input (dispersion curves) and output (S-wave velocity profiles) without specifying any explicit relationship (Hu et al., 2020; Chen et al., 2022). Here, we present "DispersioNET", a deep learning model based on convolutional neural networks (CNN) to perform joint inversion of Rayleigh wave fundamental, first and second-order mode phase velocity dispersion curves to retrieve the S-wave velocity profiles. We first describe the generation of dispersion curves from synthetic earth models, followed by a description of the architecture of DispersioNET. Finally, we illustrate the performance of DispersioNET on noise-free and noisy Rayleigh wave phase velocity dispersion curves.

## Methodology

When dealing with field data, acquired seismic data (shot gathers) are processed, followed by the generation of dispersion curves for the inversion of Rayleigh wave phase velocities. In this synthetic case study, the programs developed by Herrmann (2013) and Lehujeur et al. (2018), which incorporated the Haskell-Thomson transfer matrix method, were used to generate the dispersion curves. The S-wave velocities were sampled from a uniform distribution within suitable ranges, allowing a single CNN architecture to be trained for estimating increasing velocity with depth (IVL), intermediate low-velocity layer (LVL) and intermediate high-velocity layer models. During the generation of dispersion curves, primary (P-) wave velocities and densities were varied in tandem with S-wave velocities. However, given that Rayleigh wave phase velocity dispersion curves display a high sensitivity to S-wave velocity profiles compared to density and P-wave velocity profiles, we trained the CNN model to exclusively output S-wave velocity profiles. The S-wave velocity profiles were discretized at a uniform interval of 0.1m to enable the use of a single CNN model across varying case studies while also ensuring that the output remains independent of the number of layers. The dataset comprised 30,000 multimodal phase velocity dispersion curves and their corresponding Swave velocity profiles, of which 21000 samples were used for training (30% of training samples used for validation), and 9000 samples were used for testing. Our CNN architecture (Figure 1) has 5 convolution layers and a fully connected output layer corresponding to a total of 398,931 parameters. The weights of convolution and fully connected layers were initialized using Glorot (Glorot and Bengio, 2010) and He (He et al., 2015) initialisations, respectively. ReLU activation function was used, and batch normalisation was applied to all the convolution layers to prevent overfitting. The mean squared error (MSE) loss function was minimized with the help of Adam optimizer. The learning rate was initially set to 0.003 and reduced by half whenever the validation loss did not improve for 20 epochs.

To improve the efficacy and generalization ability of the DispersioNET architecture, we also perform training on noisy dispersion curves (Figure 2). Frequency-dependent noise was incorporated by sampling from Gaussian distributions to produce these dispersion curves. These distributions had zero mean with standard deviation decreasing linearly with increasing frequency (50 m/s and 10 m/s for the highest and lowest frequencies, respectively).

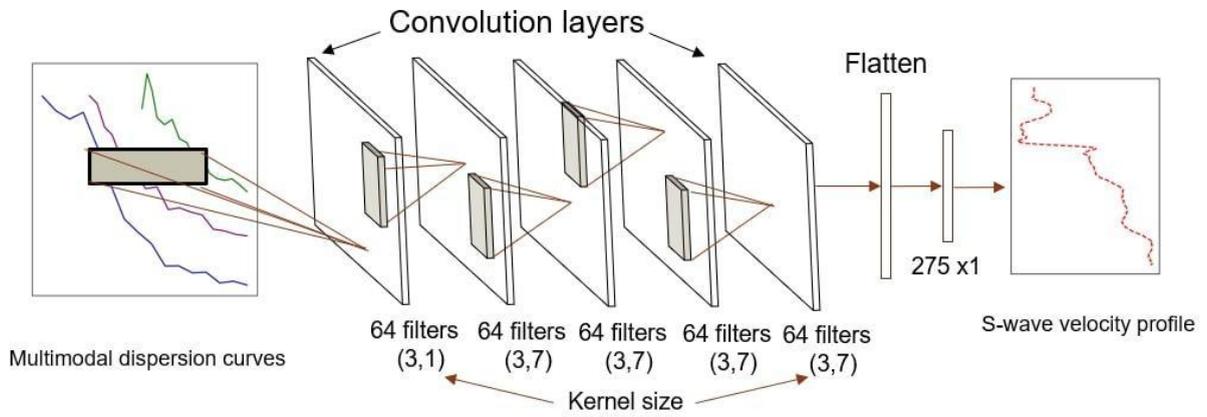

***Figure 1.*** Architecture of DispersioNET used for the joint inversion of Rayleigh wave multimode dispersion curves to retrieve S-wave velocity profiles.

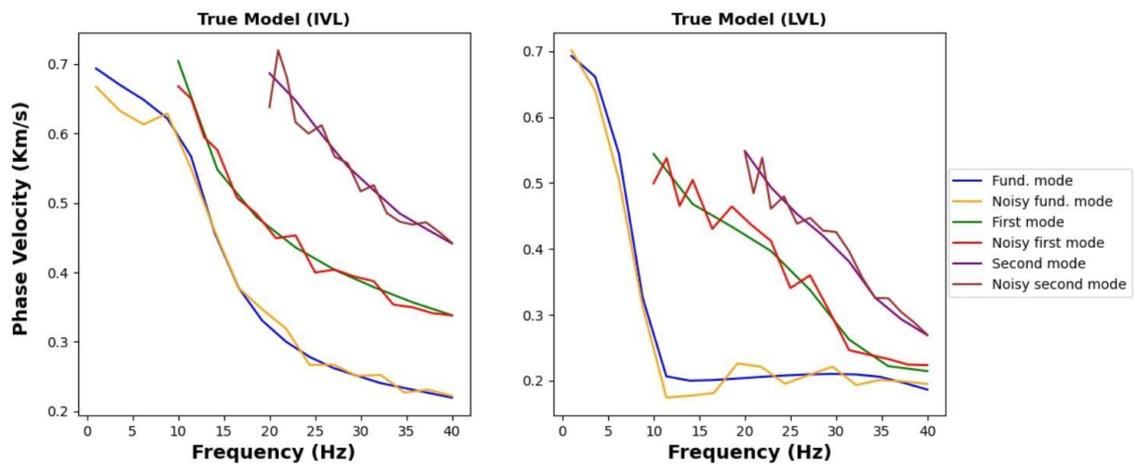

**Figure 2.** Noise-free and noisy dispersion curves corresponding to IVL and LVL models studied by Vashisth et al. (2022).

## Results and Discussion

DispersioNET was used to make predictions from the multimodal dispersion curves corresponding to the synthetic IVL and LVL models of Vashisth et al. (2022). The results (Figure 3) demonstrate a close match between the true and estimated velocity profiles, with the MSE of 0.00160 and 0.00217 for the noise-free training and test datasets, respectively. Even in the noisy case, good predictions are observed (Figure 4), albeit with some discrepancies, resulting in an MSE of 0.00371 for the training dataset and 0.00437 for the test dataset, respectively. We can observe from Figures 3 and 4 that predictions made by DispersioNET are susceptible to higher misfit as depth increases. This finding aligns with the study conducted by Vashisth et al. (2022), who reported an increase in uncertainty in the model parameters predicted for the deeper layers. Phase velocities corresponding to lower frequencies have higher uncertainties, and this contributes to the higher values of misfit at deeper layers. The DispersioNET model was trained on a significantly large set of dispersion curve and S-wave velocity profile pairs, which potentially leads to a reduction in the uncertainty of the estimated model parameters in comparison to other methods. Consequently, our future work aims to estimate not only the velocity profiles but also the uncertainties associated with them.

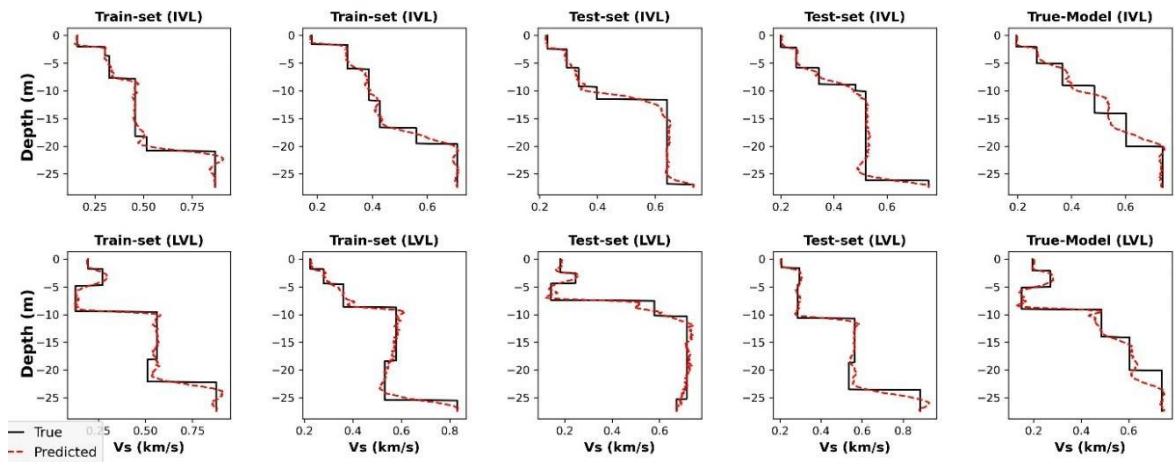

*Figure 3.* Predicted S-wave velocity profiles from noise-free train-set, test-set and true fundamental and higher order mode phase velocity dispersion curves (Figure 2) using DispersioNET (Figure 1). The true S-wave velocity profiles are plotted in black, while the inverted profiles from DispersioNET are plotted in red.

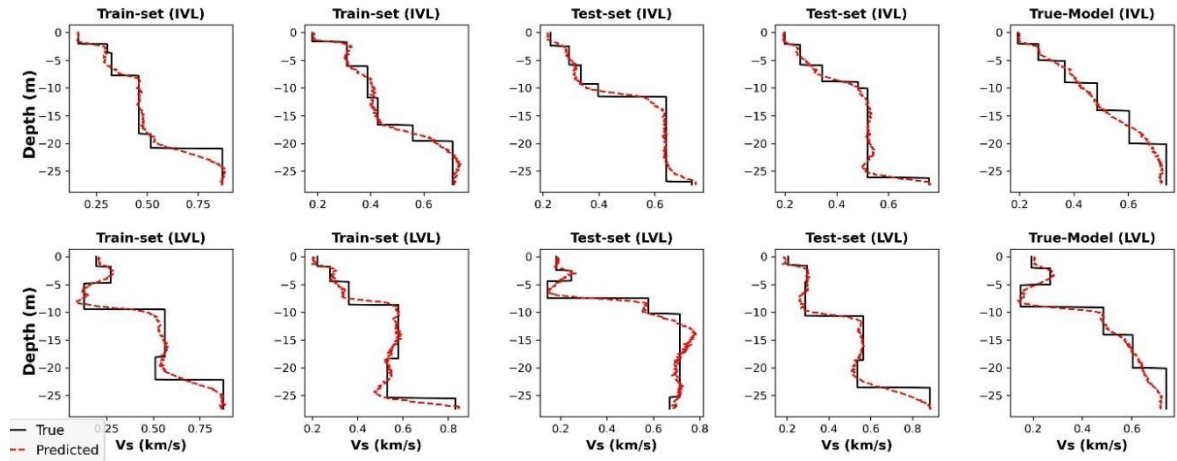

*Figure 4.* Predicted S-wave velocity profiles from noisy train-set, test-set and true fundamental and higher order mode phase velocity dispersion curves (Figure 2) using DispersioNET (Figure 1). The true S-wave velocity profiles are plotted in black, while the inverted profiles from DispersioNET are plotted in red.

## Conclusions

This study illustrates the proficiency of CNNs in implementing robust joint inversion of Rayleigh wave fundamental, first and second-order mode phase velocity dispersion curves. DispersioNET estimated S-wave velocity profiles that match well with true velocities for both noise-free and noisy case studies. DispersioNET proved capable of providing accurate predictions across a range of IVL and LVL models, with the output being independent of the number of layers in the true model.